\documentclass[proof]{pasj00}
\usepackage{ulem}
\usepackage{color}
\draft

\SetRunningHead{H.Kigure et al.}{Alfv\'en Wave Generation by Reconnection}

\title{Generation of Alfv\'en Waves by Magnetic Reconnection}
\author{Hiromitsu \textsc{Kigure}}
\email{hiromitu@kusastro.kyoto-u.ac.jp}
\affil{Kwasan and Hida Observatories, Kyoto University,Yamashina, Kyoto 607-8471}
\author{Kunio \textsc{Takahashi}}
\email{kutaka@jmastec.go.jp}
\affil{Research Institute for Global Change, JAMSTEC, 3173-25 Showa-machi, \\ Kanazawaku,
Yokohama, Kanagawa 236-0001}
\author{Kazunari \textsc{Shibata}}
\affil{Kwasan and Hida Observatories, Kyoto University,Yamashina, Kyoto 607-8471}
\author{Takaaki \textsc{Yokoyama}}
\affil{Department of Earth and Planetary Science, University of Tokyo, 7-3-1 Hongo, \\
Bunkyo-ku, Tokyo 113-0033}
\and
\author{Satoshi \textsc{Nozawa}}
\affil{Department of Science, Ibaraki University, Bunkyo 2-1-1 Mito, 
Ibaraki 310-8512}
\Received{}
\Accepted{}
\Published{}

\begin{document}

\maketitle

\begin{abstract}
In this paper, results of 2.5-dimensional magnetohydrodynamical simulations
are reported for the magnetic reconnection of non-perfectly antiparallel
magnetic fields. The magnetic field has a component perpendicular 
to the computational plane, that is, guide field. The angle $\theta$ between 
magnetic field lines in two half regions 
is a key parameter in our simulations whereas the initial distribution of 
the plasma is assumed to be simple; density and pressure are uniform 
except for the current sheet region. Alfv\'en waves 
are generated at the reconnection point and propagate along the reconnected 
field line. The energy fluxes of the Alfv\'en waves and magneto-acoustic waves (slow mode and fast mode)
generated by the magnetic reconnection are measured. Each flux shows the similar time evolution
independent of $\theta$. The percentage of the energies (time integral of energy fluxes)
carried by the Alfv\'en waves and magneto-acoustic waves to the released magnetic energy are
calculated. The Alfv\'en waves carry 38.9\%, 36.0\%, and 29.5\% of the released magnetic energy
at the maximum ($\theta=80^\circ$) in the case of $\beta=0.1$, $1$, and $20$ respectively, where
$\beta$ is the plasma $\beta$ (the ratio of gas pressure to magnetic pressure).
The magneto-acoustic waves carry 16.2\% ($\theta=70^\circ$),  25.9\% ($\theta=60^\circ$), and 75.0\%
($\theta=180^\circ$) of the energy at the maximum. Implications of these results for solar coronal
heating and acceleration of high-speed solar wind are discussed.
\end{abstract}

\section{Introduction}
The heating mechanism of the solar corona is one of the most mysterious issue 
in astrophysics (Aschwanden 2004). The structure of the solar atmosphere consists
of the photosphere, the temperature of which is about or more than six thousand
degrees, the chromosphere, the temperature of which is up to a few ten 
thousand degrees, the transition region, in which the temperature increases 
rapidly, and the corona, the temperature of which reaches a few million 
degrees. To maintain such a high temperature in the corona in spite of 
cooling by heat conduction and radiative losses, a continuous supply of 
thermal energy is necessary.

X-ray observations from early space experiments (e.g., Skylab) have shown that
the corona is not uniform and consists of many bright loops.
Poletto et al. (1975) demonstrated a correspondence between enhanced X-ray
emission and a magnetic loop. It is therefore suggested that magnetic activity
is related to the heating of the solar corona. Related to the magnetic activity,
there are two promising models for coronal heating (see reviews by Klimchuk 2006,
Erd\'elyi \& Ballai 2007 and also references therein). One is heating
by the dissipation of Alfv\'en waves
\footnote{Other types of waves (acoustic, slow-mode and fast-mode waves) are strongly
damped or reflected at the steep density and temperature gradients of chromosphere and transition region.}
that propagate in the magnetic flux tubes
(Alfv\'en 1947; Hollweg 1973, 1981, 1984, 1986; Uchida \& Kaburaki 1974; Wentzel 1974; 
McKenzie, Banaszkiewicz, \& Axford 1995; Axford et al. 1999).  
There are many proposed mechanisms for the dissipation of Alfv\'en waves,
such as mode conversion (see the next paragraph), resonant absorption (e.g., Ionson 1978; 
Poedts et al. 1989; Erd\'elyi \& Goossens 1995; Ruderman et al. 1997), phase mixing 
(e.g, Heyvaerts \& Priest 1983), or magnetohydrodynamic (MHD) turbulence (e.g., Inverarity
\& Priest 1995; Matthaeus et al. 1999).
The other is heating by many small-scale flares, i.e., nanoflares 
triggered by magnetic reconnection (Parker 1988). Observationally the occurrence frequency
of microflares and nanoflares, $N$, has been found to be $dN/dW \propto W^{-\alpha}$, where
$W$ is the total flare energy. The power-law index, $\alpha$, ranges from 1.5 to 1.8 
(Hudson 1991; Shimizu 1995; Shimojo \& Shibata 1999; Aschwanden \& 
Parnell 2002) and nanoflares can not account for coronal heating if $\alpha$ 
is less than 2. However, a value of $\alpha$ larger than 2 has also been found 
(Krucker \& Benz 1998; Parnell \& Jupp 2000). The definitive conclusion has 
not yet been obtained from observations.

As an origin of Alfv\'en waves, Kudoh \& Shibata (1999) considered a photospheric random motion propagating along
an open magnetic flux tube in the solar atomosphere, and performed 1.5-dimensional (1.5D, i.e., torsional motion
is allowed) MHD simulations for solar spicule formation and the heating of the corona. It was shown that Alfv\'en
waves transport sufficient energy flux into the corona to account for its heating, by extending the work 
by Hollweg, Jackson, \& Galloway (1982) (see also Saito, Kudoh, \& Shibata 2001).
Moriyasu et al. (2004) performed 1.5D MHD simulations of the propagation 
of nonlinear Alfv\'en waves along a closed magnetic loop including heat 
conduction and radiative cooling. They found that the corona is episodically 
heated by fast- and slow-mode MHD shocks generated by nonlinear Alfv\'en 
waves via nonlinear mode-coupling. It was also found that the time variation 
of the simulated extreme-ultraviolet (EUV) and X-ray intensities is quite similar 
to the observed one. They concluded that the observed nanoflares may not be 
a result of reconnection but may be due to nonlinear Alfv\'en waves.
Subsequently, Antolin et al. (2008) discussed the observational signatures of
the power-law indexes and coronal heating mechanisms (Alfv\'en waves and nanoflares)
by using 1.5D MHD model of Moriyasu et al. (2004). They found that Alfv\'en heating
and nanoflare heating exhibit different power-law indexes (see also Antolin \& Shibata 2010).

However, since Alfv\'en waves can be generated by magnetic reconnection unless the
reconnection takes place in perfectly antiparallel magnetic fields, the nanoflare
model may not be very different from the Alfv\'en wave model. Yokoyama \& Shibata (1995)
modelled jets (X-ray or EUV jets and serges observed with H$\alpha$ in the chromosphere)
by performing a resistive two-dimensional MHD simulation of the
magnetic reconnection occurring in the current sheet between emerging magnetic flux and overlying
pre-existing coronal magnetic fields. Recent {\it Hinode} (Kosugi et al. 2007) observation revealed
that jets are ubiquitous in the chromosphere (Shibata et al. 2007). Nishizuka et al. (2008) proved
that the jets on the basis of this model are quantitatively corresponding to the multiwavelength jets
(see also Cirtain et al. 2007 and Liu et al. 2009) observed with {\it Hinode} and {\it TRACE}
(Handy et al. 1999).  Yokoyama (1998) found that the ratio of energy of Alfv\'en waves
to the energy released by the magnetic reconnection (the model of Yokoyama \& Shibata 1995) is nearly
equal to 3\%. In this model, there is a shear between emerging magnetic fields and coronal magnetic fields,
kinks are produced by the magnetic reconnection and propagate away as Alfv\'en waves. 

Takeuchi \& Shibata (2001) performed 2.5-dimensional (2.5D) MHD simulations of the photospheric
magnetic reconnection caused by convection, and found that the energy flux of Alfv\'en waves
\footnote{The perpendicular magnetic field injected by magnetic reconnection propagates
as Alfv\'en waves.} is enough to explain both coronal heating and spicule production. The generation of
Alfv\'en waves through magnetic reconnection has been also discussed by Parker (1991), Axford et al. (1999),
and Sturrock (1999) in the context of coronal hole heating and solar wind acceleration.

The upward acoustic waves in the flux tube are expected to be transported as other waves
(e.g., slow-mode waves) after the formation of the shocks.
Slow-mode waves are also thought to contribute to coronal heating though the 
contribution is only to the relatively low corona because of their 
compressibility. Takeuchi \& Shibata (2001) measured the energy flux of 
slow-mode waves and found that it was ten times larger than that of Alfv\'en 
waves. Suzuki (2002) discussed the possibility of coronal heating and the 
acceleration of the low-speed solar wind by slow-mode waves. 
Suzuki (2004) developed his study by including fast-mode waves (linearly 
polarized Alfv\'en waves) and showed that slow-mode waves contribute to the 
heating of the low corona and the acceleration of the low-speed solar wind, 
while linearly polarized Alfv\'en waves contribute to the heating of the outer 
corona and the acceleration of the high-speed solar wind. 

As the mention above, waves are created by torsional motions as recently observed (e.g., Bonet et al. 2008)
in the photosphere or magnetic reconnections. They propagate into the corona and disipate their energy through
linear and non-linear mechanisms. In particular, Alfv\'en waves can be generated by such drivers, for the first
time, by Jess et al. (2009). From {\it Hinode} data, De Pontieu et al. (2007) estimated the energy flux carried by
transversal oscillations generated by spicules and compared with radiative MHD simulations by more realistically
extending Kudoh \& Shibata (1999). They indicated that the calculated energy flux is enough to heat the quiet corona
and to accelerate the high-speed solar wind. Okamoto et al. (2007) also estimated the energy flux to be $2\times 10^6$
ergs cm$^{-2}$ s$^{-1}$ propagating on coronal magnetic fields. These reports (see also Tomczyk et al. 2007), however, have
considerable argument about what is Alfv\'en waves. Erd\'elyi and Fedun (2007) and Van Doorsselaere et al. (2008) argued that
these oscillations were likely to be kink oscillations from  observed behavior (see also Goossens et al. 2009; Taroyan \& Erd\'elyi 2009).

It is thought that the magnetic reconnection causes the solar flare. Particle 
acceleration takes place associated with solar flare. The mechanism of the 
particle acceleration, however, is not made clear. Alfv\'en waves 
could contribute to the acceleration of ions through cyclotron resonance (see, 
e.g., Miller 2000, and references therein). The generation of Alfv\'en waves 
by the magnetic reconnection is a very interesting research topic also from 
such a point of view.

In this paper, we present the results of 2.5D MHD simulations of 
the Alfv\'en wave generation by the reconnection of non-perfectly 
antiparallel magnetic fields. The initial magnetic fields have a shear, i.e., 
the magnetic field has a component perpendicular to the computational plane. 
This component can not contribute to the reconnection and is so-called guide 
field. 
The magnetic reconnection in this geometry was analytically studied by 
Petschek \& Thorne (1967).
The angle between the magnetic field lines in two half regions is a parameter, 
$\theta$, while the initial distribution of the plasma is assumed to be 
simple. The energy fluxes of Alfv\'en waves and magneto-acoustic waves are measured. 
In section 2 we describe the numerical method and model. In section 3 we show the 
results of the simulations. Finally a discussion and summary are given in 
section 4.
\section{Numerical Simulations}

\subsection{Assumptions and Basic Equations}
In this paper, we assume that the distributions of the physical quantities 
are not dependent on the $z$-coordinate but that the vector fields have the 
$z$-component (i.e., 2.5D approximation).
We solve the following resistive MHD equations numerically: 
\begin{eqnarray}
\frac{\partial\rho}{\partial t}+({\boldsymbol v}\cdot\nabla)\rho
=-\rho(\nabla\cdot{\boldsymbol v}),
\end{eqnarray}
\begin{eqnarray}
\frac{\partial{\boldsymbol v}}{\partial t}+({\boldsymbol v}\cdot\nabla){\boldsymbol v}
=-\frac{1}{\rho}\nabla p+\frac{1}{4\pi\rho}\left({\boldsymbol j}\times{\boldsymbol B}\right),
\end{eqnarray}
\begin{eqnarray}
\frac{\partial p}{\partial t}+({\boldsymbol v}\cdot\nabla)p
=-\gamma p(\nabla\cdot{\boldsymbol v}) +(\gamma-1)\frac{\eta}{4\pi}|{\boldsymbol j}|^2,
\end{eqnarray}
\begin{eqnarray}
\frac{\partial{\boldsymbol B}}{\partial t} = - \nabla \times {\boldsymbol E},
\end{eqnarray}
\begin{eqnarray}
{\boldsymbol E} = \eta {\boldsymbol j} - {\boldsymbol v} \times {\boldsymbol B},
\end{eqnarray}
and
\begin{eqnarray}
{\boldsymbol j} = \nabla \times {\boldsymbol B},
\end{eqnarray}
where $\rho, p,{\boldsymbol v}$ are the density, pressure, and velocity 
of the gas, ${\boldsymbol B},{\boldsymbol E},{\boldsymbol j}$ are the magnetic field, electric field, and  
current density, and $\gamma$,  $\eta$ represent the ratio of specific heats and electric resistivity.

\subsection{Initial Conditions and Boundary Conditions}
We consider the situation that the physical quantities are uniform far from 
the current sheet. Their typical quantities are $\rho_0=1$, $p_0=1/\gamma$, and
$B_0=|{\boldsymbol B}_0|=\sqrt{8\pi/(\beta\gamma)}$, where ${\boldsymbol B}_0$ is
the initial magnetic field. $\beta = 8\pi p_0/{B_0}^2$ is the plasma $\beta$
(the ratio of gas pressure to magnetic pressure). Initially the magnetic pressure
gradient force balances with the gas pressure gradient force. The sound speed,
$C_\mathrm{S}=\sqrt{\gamma p/\rho}$, is unity in the whole domain of the
simulation box.

The initial distributions of the physical quantities are as follows: 
\begin{eqnarray}
B_y = B_0 &\tanh\left(2\frac{x}{L_0}\right)& \cos\phi, \\ 
B_z = B_0 &\tanh\left|2\frac{x}{L_0}\right|& \sin\phi,
\end{eqnarray}
\begin{eqnarray}
B_x &=& v_x = v_y = v_z = 0, \\
p &=& p_0 \left( 1 + \frac{1}{\beta} \right) - \frac{B_y^2 + B_z^2}{8 \pi},
\end{eqnarray}
and
\begin{eqnarray}
\rho &=& \gamma p,
\end{eqnarray}
where
\begin{eqnarray}
\phi &=& 90^\circ - \frac{\theta}{2},
\end{eqnarray}
$L_0$ is the thickness of the initial current sheet. In this study, we have adopted $\gamma=5/3$
and $L_0=1$. 

\begin{figure}
\begin{center}
\FigureFile(100mm,100mm){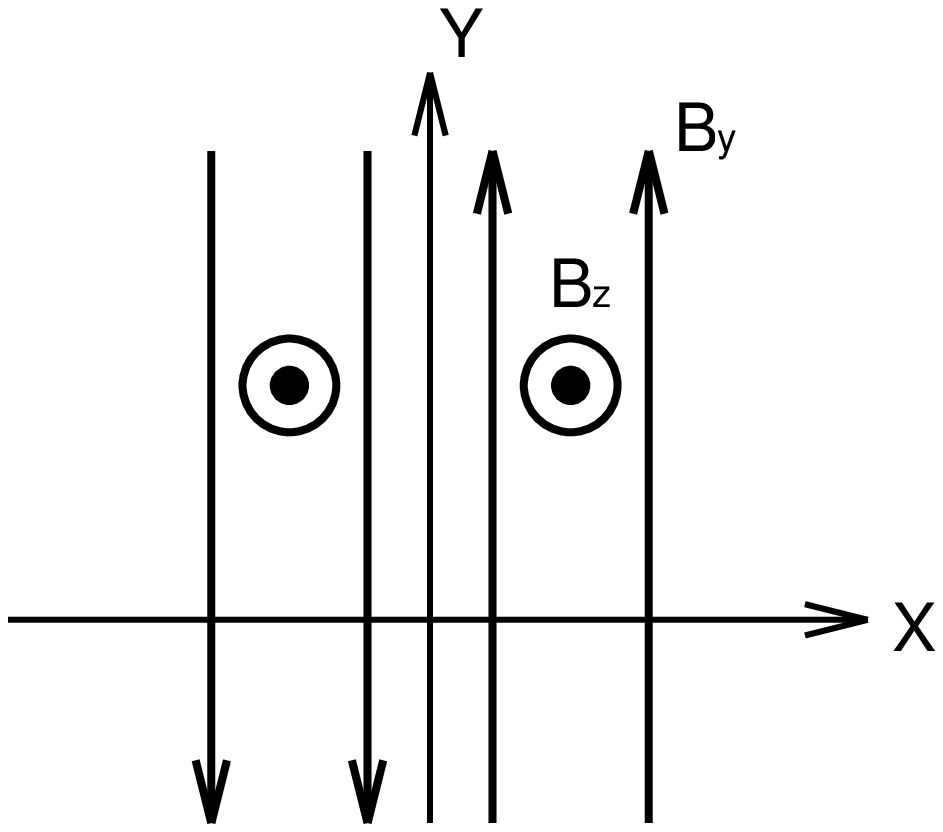}
\FigureFile(100mm,100mm){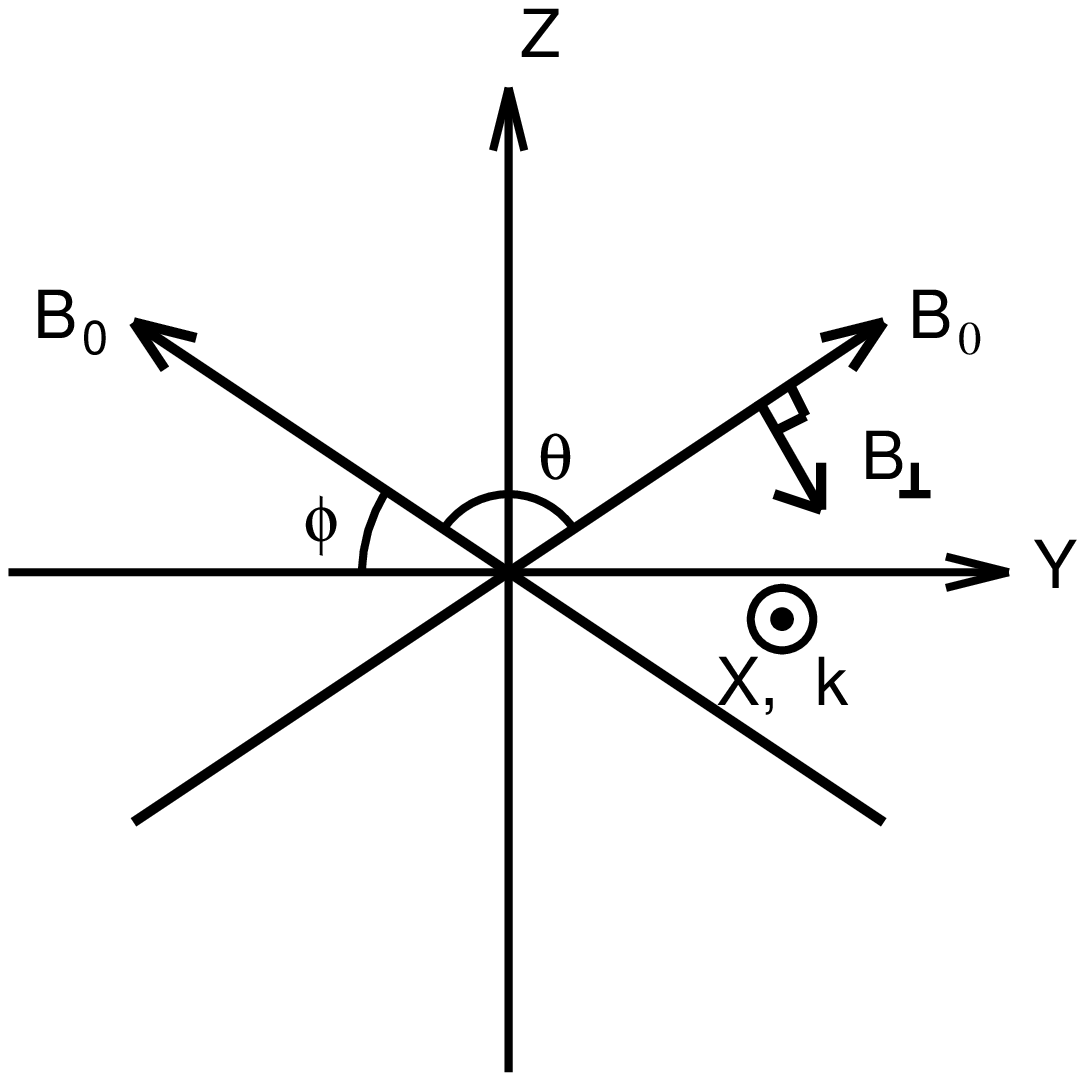}
\end{center}
\caption{Coordinate systems ($xy$-plane and $yz$-plane) of the initial magnetic field 
configuration in this study.}
\label{FIG01}
\end{figure}

Figure \ref{FIG01} shows the coordinate system ($xy$-plane and $yz$-plane),
the initial magnetic field configuration, and the definition of $\theta$ and
$\phi$ (the directions of the wave vector and $B_\perp$ discussed in section
\ref{Sec:Time-Evo} are also displayed).  When $\theta$ is equal to $180^\circ$
the magnetic fields are perfectly antiparallel on both sides of a line where $x = 0$. 
We investigated  the cases where $\theta$ is from $180^\circ$ to $10^\circ$. 
The value of the plasma $\beta$ is also a variable parameter and equal to 
0.1, 1, or 20. All physical quantities are normalized by their typical values.
This means that the velocity is normalized by the initial sound speed,
$C_\mathrm{S0}=\sqrt{\gamma p_0/\rho_0}$. Time is normalized by $t_0=L_0/C_\mathrm{S0}$.

The number of grid points in the simulations is $603 \times 1003$. The grid 
spacing is uniform within $-5L_0 \le x \le 5L_0$ and $-20L_0 \le y \le 20L_0$. 
The constant grid spacing in the $x$-direction is equal to $0.025L_0$ and that in 
the $y$-direction is $0.05L_0$. The grid spacing in the non-uniform region is 
slowly stretched by an increment of 2 \% at each grid step (e.g., $|\Delta x_{i+1}|=1.02|\Delta x_i|$).
The size of the computational domain is $-13.1L_0 \le x \le 13.1L_0$,
$-36.3L_0 \le y \le 36.3L_0$.

At all the boundaries of the simulation box the periodic boundary condition is 
assumed. The total energy is therefore conserved. It can be said that the 
decrease of the magnetic energy in the simulation box is equal to the energy 
released by the magnetic reconnection.

\subsection{Resistivity Model and Numerical Method}
In order to initiate the magnetic reconnection, we assume a localized 
resistivity near the origin, $\left( x, y \right) = \left( 0, 0 \right)$, 
as 
\begin{eqnarray}
\eta = \left\{
\begin{array}{ll}
\eta_0 \left[ 2 \left( r/r_\eta \right)^3 - 3 \left( r/r_\eta \right)^2 
+ 1 \right], &\left( r \leq r_\eta \right), \\
0, &\left( r > r_\eta \right), \\
\end{array} \right.
\end{eqnarray}
where $r = \left( x^2 + y^2 \right)^{1/2}$, $r_\eta = L_0$, and $\eta_0 = 0.1C_\mathrm{S0}L_0$.
The magnetic Reynolds number is $R_\mathrm{m} \equiv v_\mathrm{A0} L_y / \eta_0$,
where $v_\mathrm{A0}$ is the initial Alfv\'en velocity outside the current sheet and $L_y
(=y_\mathrm{max}-y_\mathrm{min}=72.6 L_0)$ is the size of the computational domain in the $y$-direction.
The values of the plasma $\beta$ and $R_\mathrm{m}$ in our simulations are summarized in Table 1.
We here mention that this non-dimensional parameter, $R_\mathrm{m}$, is much smaller than
that of typical solar corona plasma ($R_\mathrm{m} \sim 10^{14}$). However the numerical treatment of
$R_\mathrm{m} \sim 10^{14}$ is very difficult in current-day computer resources.

The numerical computations were carried out by the Rational CIP (Cubic interpolated profile) method
(Yabe \& Aoki 1991; Xiao, Yabe \& Ito 1996) combined with the MOC-CT method (Evans \& Hawley 1988; Stone \& Norman 1992).
The magnetic induction equation was solved by the MOC-CT and the other equations were solved by the CIP
(e.g., Kudoh, Matsumoto \& Shibata 1998; Kigure \& Shibata 2005; Takahashi et al. 2009).

\section{Results of Numerical Simulations}

\subsection{Time Evolution}\label{Sec:Time-Evo}
\begin{figure}
\begin{center}
\FigureFile(160mm,160mm){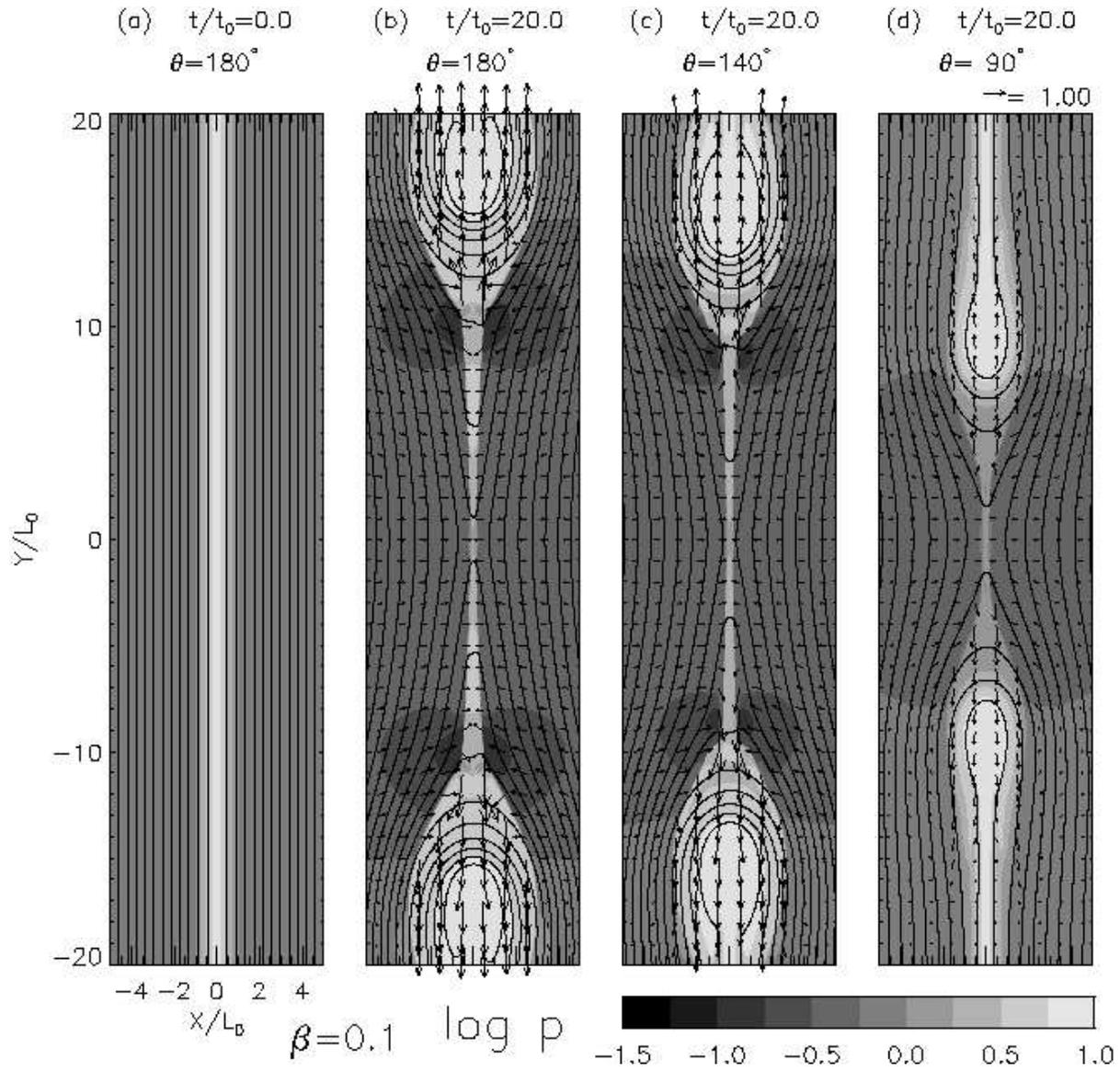}
\end{center}
\caption{Distribution of the logarithmic pressure in the case of
(a) $\theta=180^\circ$ at initial state, (b)  $\theta=180^\circ$ at $t/t_0=20.0$, (c) $\theta=140^\circ$
at $t/t_0=20.0$ and (d) $\theta=90^\circ$ at $t/t_0=20.0$. The plasma $\beta$ is 0.1. The black lines show
the magnetic field lines, and the arrows show the velocity.}
\label{FIG02}
\end{figure}

First we describe the time evolution of the system only in the case of 
plasma $\beta = 0.1$. Figures \ref{FIG02}a and \ref{FIG02}b
show the distribution of the logarithmic pressure in the case of $\theta = 180^\circ$.
The magnetic fields reconnect in the diffusion region, which is around $(x,y) = (0,0)$
and the magnetic energy is converted to thermal energy. The plasma is accelerated in
the $y$-direction by the magnetic tension of the reconnected
magnetic field. The diffusion region is localized so that the Petschek-type 
reconnection takes place. The slow-mode MHD shock is therefore formed, i.e.,
the gas pressure increases and the magnetic pressure decreases behind the 
shock front. The velocity of the reconnection outflow is the order of 
Alfv\'en velocity. These are equivalent to the magnetic energy being 
converted to thermal and kinetic energy through the shock front.

\begin{figure}
\begin{center}
\FigureFile(100mm,100mm){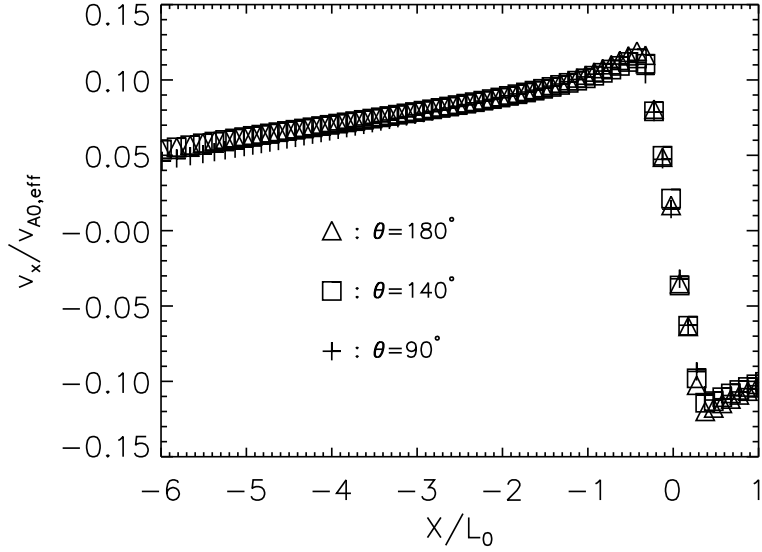}
\end{center}
\caption{The distribution of the $x$-component of the velocity normalized by 
$v_\mathrm{A0} \cos \left( 90^\circ - \theta/2\right) = v_\mathrm{A0,eff}$ on a line 
where $y = 0$ at $t/t_0 = 20$. The triangles display the results of the 
$\theta = 180^\circ$ case, the squares display the results of the 
$\theta = 140^\circ$ case, and the plus signs display the results of the 
$\theta = 90^\circ$ case. The plasma $\beta$ is 0.1. The inflow velocities, 
$v_{x}$ in this case, are roughly proportional to $\cos \left( 90^\circ - \theta/2\right)$. 
This can be interpreted as $v_{x} \propto v_\mathrm{A0,eff}$, 
where $v_\mathrm{A0,eff}$ is the Alfv\'en velocity in the computational plane.}
\label{FIG03}
\end{figure}

Figures \ref{FIG02}c and \ref{FIG02}d show the distribution in the case of 
$\theta = 140^\circ$ and $\theta = 90^\circ$. The $y$-components of the 
initial magnetic field in these cases are less than that in the 
$\theta = 180^\circ$ case (the difference in the number of the contour 
indicates this situation) while the $z$-components are not equal to zero in 
these cases. 
Figure \ref{FIG03} shows the distribution of the $x$-component of the 
velocity normalized by $v_\mathrm{A0} \cos \left( 90^\circ - \theta/2\right) = 
v_\mathrm{A0,eff}$ on a line where $y = 0$ at $t/t_0 = 20$. The triangles display the 
results of the $\theta = 180^\circ$ case, the squares display the 
results of the $\theta = 140^\circ$ case, and the plus signs the $\theta 
= 90^\circ$ case. The inflow velocities $v_\mathrm{in}$, especially near the current 
sheet $(|x|/L_0 < 2)$, are proportional to $v_\mathrm{A0,eff}$.

From the linearized MHD equations, the disturbance of velocity and magnetic 
field due to the Alfv\'en wave is perpendicular to the initial (non-perturbed) 
magnetic field, ${\boldsymbol B}_0$, and the wave vector, 
${\boldsymbol k}$ (i.e., parallel to ${\boldsymbol k}
\times{\boldsymbol B}_0$). In our case, the initial magnetic field 
is on the $yz$-plane. For simplicity, we assume that the wave vector is 
approximated to be parallel to the $x$ (or $-x$) direction (the Alfv\'en wave 
front is on the $yz$-plane). In this case, the perturbed components of 
velocity and magnetic field are on the $yz$-plane (no $x$-component) and 
perpendicular to the initial field. Hereafter, we call these 
velocity and magnetic field disturbances due to the Alfv\'en wave as 
$v_\perp$ and $B_\perp$. In this case, the perturbed components are described as
$v_\perp =v_y \sin\phi - v_z \cos\phi$ and $B_\perp =B_y \sin\phi - B_z \cos\phi$
(see Figure \ref{FIG01}). Note that the non-linear effect of shock
propagating through the magnetic reconnection is included. However, it is difficult
to remove the non-linear effect from  these components. Therefore, this assumption would
be acceptable for our first attempt to study the magnetic disturbance of non-perfectly
antiparallel magnetic reconnection.

\begin{figure}
\begin{center}
\FigureFile(160mm,160mm){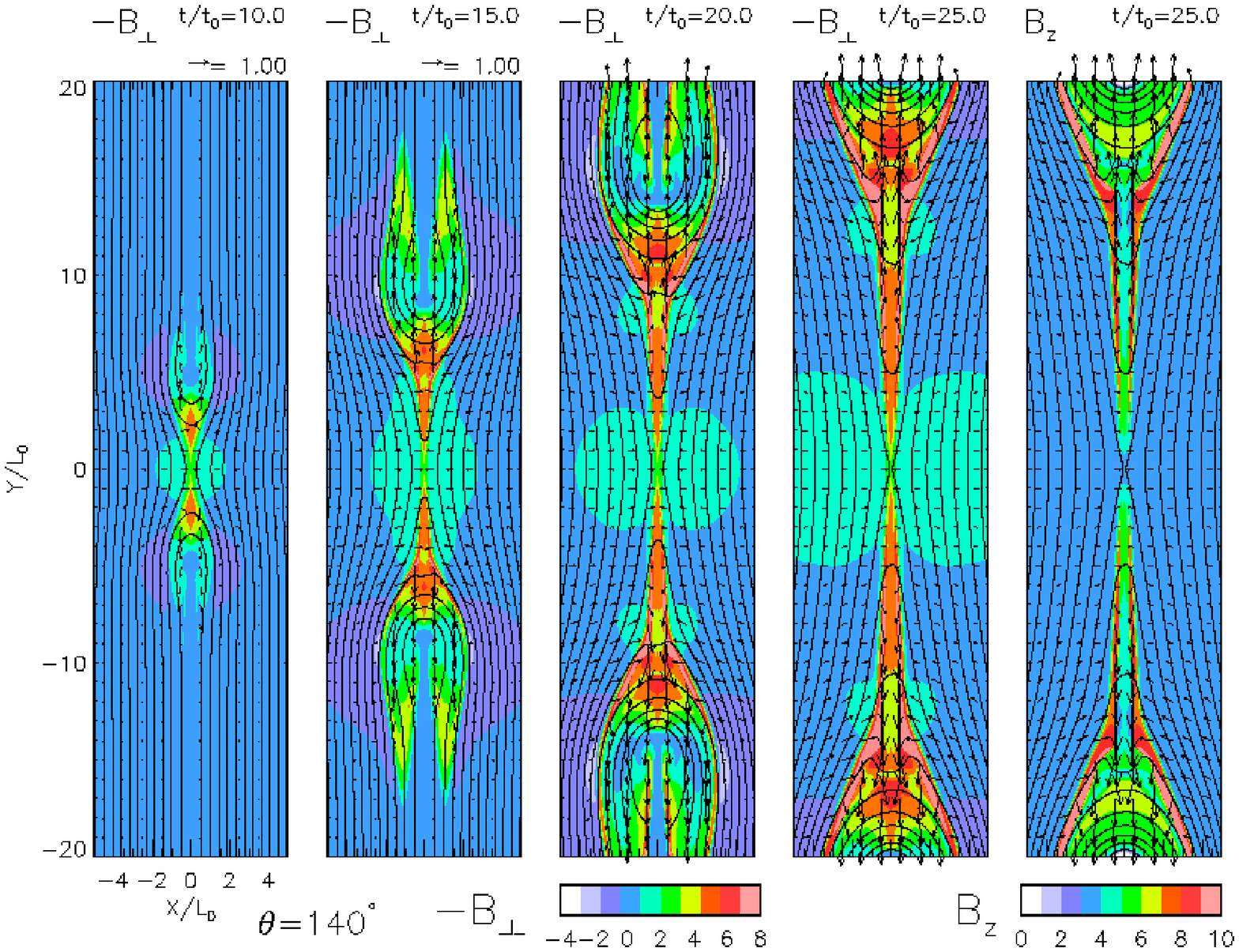}
\end{center}
\caption{The left four panels show the time evolution of the magnetic field 
component perpendicular to the initial field, $-B_\perp$, in the case of 
$\theta=140^\circ$. The right panel shows the $B_z$ distribution at $t/t_0=25$. 
The plasma $\beta$ is 0.1. The black lines show the 
magnetic field lines, and the arrows show the velocity.}
\label{FIG04}
\end{figure}

\begin{figure}
\begin{center}
\FigureFile(160mm,160mm){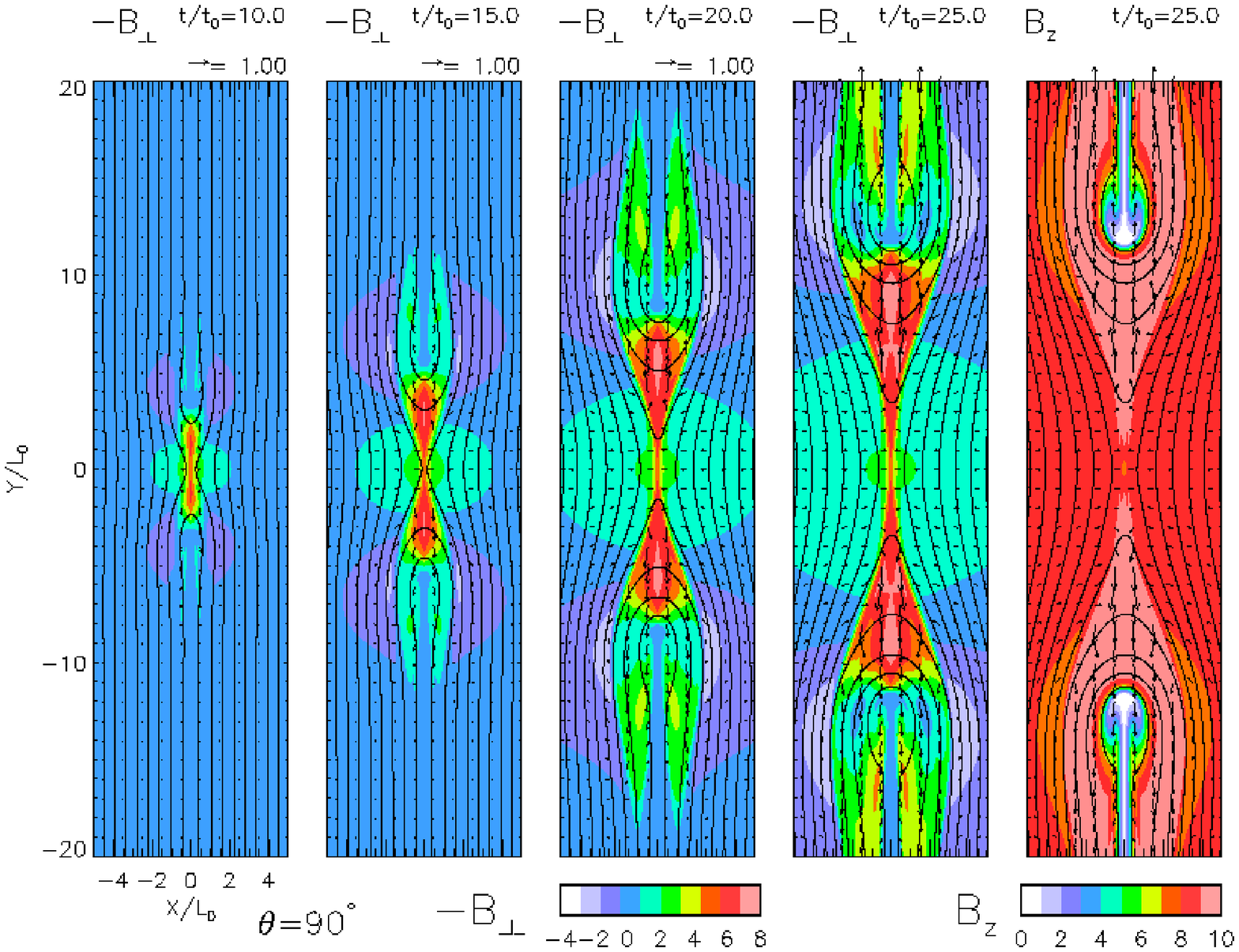}
\end{center}
\caption{The left four panels show the time evolution of the magnetic field 
component perpendicular to the initial field, $-B_\perp$, in the case of 
$\theta=90^\circ$. The right panel shows the $B_z$ distribution at $t/t_0=25$. 
The plasma $\beta$ is 0.1. The black lines show the 
magnetic field lines, and the arrows show the velocity.}
\label{FIG05}
\end{figure}

In the case where there exists a magnetic field component perpendicular to 
the $xy$-plane, the field component perpendicular to the initial field, 
$B_\perp$, is generated as a result of the reconnection although $B_\perp$ 
is not generated in the perfectly antiparallel reconnection (in the perfectly 
antiparallel reconnection case, $B_\perp$ means $B_z$). The generated 
$B_\perp$ propagates as Alfv\'en waves along field lines. The left four
panels of Figures \ref{FIG04} and \ref{FIG05} show the time evolutions of $-B_\perp$
in the cases of $\theta = 140^\circ$ and $\theta = 90^\circ$ respectively. 
The contour lines show the magnetic field lines and the arrows show the 
velocity.
The right panels of Figures \ref{FIG04} and \ref{FIG05} show the $B_z$ 
distribution at $t/t_0=25$. Following a certain magnetic field line, e.g., from 
the point $(x/L_0,y/L_0)=(-3,20)$ in the $\theta=140^\circ$ case, $B_z$ once 
increases and then decreases with keeping its sign. 
This feature is consistent with the analytical solution in Petschek \& 
Thorne (1967) (see Figure 3 in that paper). The point where $B_z$ 
increases corresponds to Alfv\'en wave and the point where $B_z$ 
decreases corresponds to slow-mode MHD shock.

\subsection{Energy Fluxes of Alfv\'en Waves and Magneto-acoustic Waves}
In this study, we focus on the disturbance energy fluxes carried by the Alfv\'en waves ($F_\mathrm{Alfven}$) and 
magneto-acoustic waves ($F_\mathrm{Sound}$) generated by the magnetic reconnection. These fluxes are measured on a line where
$y/L_0=\pm10$.
\begin{eqnarray}
F_\mathrm{Alfven} &=&  \pm \frac{1}{4\pi L_x} \int_{x_\mathrm{min}}^{x_\mathrm{max}} - B_\perp v_\perp B_{\parallel y} \ dx, \\
F_\mathrm{Sound} &=& \pm \frac{1}{L_x} \int_{x_\mathrm{min}}^{x_\mathrm{max}} \delta p \ v_{\parallel y} \ dx,
\end{eqnarray}
where $\delta$ means the difference from the initial value, $L_x$($=x_\mathrm{max}-x_\mathrm{min}=26.2L_0$)
is the size of the computational domain in the $x$-direction. $v_{\parallel y}$ ($B_{\parallel y}$) is the
$y$-component of velocity (magnetic field) parallel to the initial magnetic field, ${\boldsymbol B_0}$.
$v_{\parallel y}$ and $B_{\parallel y}$ are described as
\begin{eqnarray}
v_{\parallel y}&=&\frac{v_y{B_{y0}}^2+v_zB_{y0}B_{z0}}{{B_{y0}}^2+{B_{z0}}^2}, \\
B_{\parallel y}&=&\frac{B_y{B_{y0}}^2+B_zB_{y0}B_{z0}}{{B_{y0}}^2+{B_{z0}}^2}.
\end{eqnarray}
Here, $B_{y0}$ ($B_{z0}$) is the $y(z)$-component of the initial magnetic field.
The plus sign is on the $y/L_0=10$ line and the minus sign on the $y/L_0=-10$ line. We also measure the time
integral of each flux:
\begin{eqnarray}
E_\mathrm{Alfven} &=& L_x \int_{0}^{t} F_\mathrm{Alfven}\ dt, \\
E_\mathrm{Sound} &=& L_x \int_{0}^{t} F_\mathrm{Sound}\ dt.
\end{eqnarray}
$F_\mathrm{Alfven}$ and $F_\mathrm{Sound}$ are normalized by $\rho_0 C_\mathrm{S0}^3$.
$E_\mathrm{Alfven}$ and $E_\mathrm{Sound}$ are normalized by $\rho_0 C_\mathrm{S0}^2 L_0^2$.
The origin of the energies transported by the Alfv\'en waves and
magneto-acoustic waves is the magnetic energy released by the reconnection so that the 
magnetic energy in the simulation domain is also measured.

Figures \ref{FIG06}a-c show the time evolutions of the released magnetic 
energy ($\Delta E_\mathrm{mg} \equiv E_\mathrm{mg0} -  E_\mathrm{mg}$) in the
simulation domain, where $E_\mathrm{mg}$ is defined by
\begin{eqnarray}
E_\mathrm{mg} =  \frac{1}{8\pi} \int_{x_\mathrm{min}}^{x_\mathrm{max}} \int_{y_\mathrm{min}}^{y_\mathrm{max}} |{\boldsymbol B}|^2 \ dx \ dy,
\end{eqnarray}
and $E_\mathrm{mg}$ is normalized by $\rho_0 C_\mathrm{S0}^2 L_0^2$. Here, $E_\mathrm{mg0}$ is the
initial magnetic energy in the simulation domain.
It is clear that the amount of the released magnetic energy decreases as the angle
of magnetic shear, $\theta$, decreases and the plasma $\beta$ increases. In the final
evolutional stage of each calculation, the magnetic energy in the simulation domain decreases linearly 
with time. Figure \ref{FIG06}d shows the value of $\mid dE_\mathrm{mg}/dt 
\mid$ in the case of $\theta = 180^\circ, 140^\circ$ and $90^\circ$. 
The magnetic energy released by the magnetic reconnection is equivalent to the 
Poynting flux entering into the reconnection region.
Since the reconnection inflow, $v_\mathrm{in}$, is expressed as $v_\mathrm{in} 
= \epsilon v_\mathrm{A} \propto B$ in the case of Petschek-type reconnection, where 
$\epsilon$ is the reconnection rate and roughly independent of $B$ (e.g., 
Magara et al. 1996, Yokoyama \& Shibata 1997), 
\begin{eqnarray}
-\frac{dE_\mathrm{mg}}{dt} \sim 2 \frac{B^2}{4\pi}l v_\mathrm{in} \propto B^3 \propto 
\beta^{-3/2}\label{EQEmgbeta},
\end{eqnarray}
where $l$ is the size of reconnection region (Tanuma et al. 1999). 
The solid line in Figure \ref{FIG06}d shows a line where 
$\mid dE_\mathrm{mg}/dt \mid \propto \beta^{-3/2}$. It is clear from Figure 
\ref{FIG06}d that the Equation (\ref{EQEmgbeta}) is almost satisfied 
independent of $\theta$.

Figures \ref{FIG07}a and \ref{FIG07}b show the time evolutions of the energy fluxes 
carried by the Alfv\'en waves and magneto-acoustic waves,
$F_\mathrm{Alfven}$ and $F_\mathrm{Sound}$, in the case of $\beta = 0.1$.
If $\theta = 180^\circ$, $F_\mathrm{Alfven}$ is always zero
because the perturbation perpendicular to the initial field is not generated. 
The time evolutions of the magneto-acoustic wave
flux are similar in each case except when $\theta \leq 60^\circ$. Those of the Alfv\'en wave
flux are also similar except for the cases where $\theta=180^\circ$ and $\theta \leq 60^\circ$.
The magneto-acoustic and Alfv\'en wave
fluxes decrease after reaching the peak value. 

Figure \ref{FIG07}c and \ref{FIG07}d are the case of $\beta=1$.
The time evolutions of $F_\mathrm{Alfven}$ and $F_\mathrm{Sound}$ in this case are
similar in each $\theta$ case respectively and similar to those in the $\beta = 0.1$ case. 
There is a following tendency; $F_\mathrm{Alfven}$ at the late stage once
increases and then decreases as $\theta$ decreases, while $F_\mathrm{Sound}$
at the late stage decreases monotonically. Figures \ref{FIG07}e and \ref{FIG07}f show the results for the
case of $\beta=20$. Though the same features are seen,  $F_\mathrm{Alfven}$ and $F_\mathrm{Sound}$ are
different by more than one order of magnitude when compared with the $\beta=1$ case (the time at the late
stage of the $\beta=20$ case is about 3.5 times larger than that of the $\beta=1$ case, i.e., the time
integral is made about 3.5 times longer).

\section{Discussions and Summary}
Figure \ref{FIG08} shows the ratio (percentage) of the energies carried by 
the Alfv\'en waves and magneto-acoustic waves ($E_\mathrm{Alfven}$ and $E_\mathrm{Sound}$) to the 
released magnetic energy at the final stage of simulation. In the case of
$\theta=180^\circ$, $E_\mathrm{Alfven}$ is equal to 0 independent of the plasma $\beta$
because the magnetic field component perpendicular to the initial field is not generated by the
reconnection. As $\theta$ decreases the percentage of the energy carried by the Alfv\'en waves
increases up to 38.9\%, 36.0\%, and 29.5\% in the case of $\beta=0.1$, $1$, and $20$ 
respectively, and then decreases. The percentage is maximum when $\theta = 80^\circ$.
The percentage of the energy carried by the magneto-acoustic waves is
roughly constant when $\theta$ is relatively large in the cases of $\beta=0.1$ and $1$. In the case of
$\beta=20$, it gradually decreases in decreasing $\theta$. The maximum values are 16.2\%, 25.9\%, and 75.0\%
in the case of $\beta=0.1$, $1$, and $20$ respectively.

While the percentage of the energy carried by the Alfv\'en waves is almost 
independent of $\beta$, that of the energy carried by the magneto-acoustic waves
changes by some factor. The ratio of the energy carried by the magneto-acoustic waves
to the released magnetic energy becomes larger as $\beta$ becomes larger in $\theta\geq 40$.
This means that the significant part of the energy released by the magnetic reconnection is transported
as a perturbation of gas pressure in a high-$\beta$ plasma case.

This result can give a suggestion to the study of high-$\beta$ plasma 
astrophysical objects, e.g., the accretion disk. The accretion disk is 
thought to be weakly magnetized. The magnetic reconnection is induced by 
the magnetorotational instability (e.g., Sano \& Inutsuka 2001, Machida \& 
Matsumoto 2003). Sano \& Inutsuka (2001) showed that the heating rate is 
strongly related to the turbulent shear stress, which determines the 
efficiency of angular momentum transport. Therefore, the study of the 
magnetic reconnection in a high-$\beta$ plasma is important.

The total non-radiative energy input to the solar coronal hole was estimated 
at $5 \times 10^5$ ergs cm$^{-2}$ s$^{-1}$ by Withbroe (1988). For the 
acceleration of high-speed solar wind, some $1 \times 10^5$ ergs cm$^{-2}$ 
s$^{-1}$ is required to be deposited at distances of several solar radii 
(see, e.g., Parker 1991, and references therein). 
If the solar wind is accelerated by the energy flux of Alfv\'en waves,  
this means that 20\% of the 
energy released by reconnection events in the solar corona is transferred as a 
form of Alfv\'en wave. Our results show that the energy larger than the 
required can be carried by the Alfv\'en wave independent of $\beta$ around 
the parametric region of $60^\circ \leq \theta \leq 110^\circ$. These 
energies are converted to thermal energy through the dissipation of Alfv\'en 
waves.

\begin{figure}
\begin{center}
\FigureFile(160mm,160mm){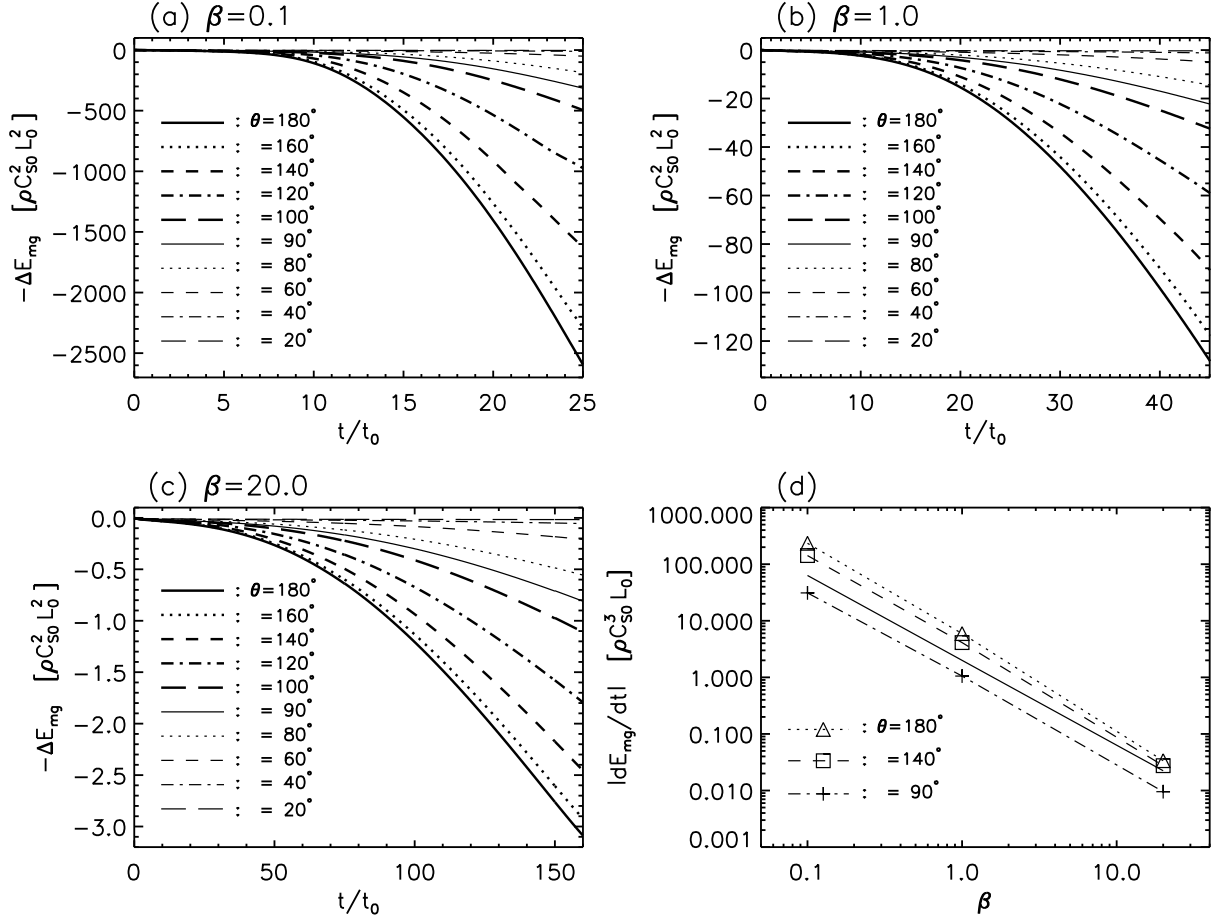}
\end{center}
\caption{(a)-(c) Time evolution of the released magnetic energy in the 
computational domain in each $\theta$ case. The plasma $\beta$ are
(a) 0.1, (b) 1, and (c) 20. (d) The dependence of the magnetic energy release rate
in the late stage on $\beta$ in each $\theta$ case. The solid line shows a line
where $\mid dE_\mathrm{mg}/dt \mid \propto \beta^{-3/2}$. The magnetic energy
release rate is proportional to $\beta^{-3/2}$, which is consistent with the
theoretical prediction, independent of $\theta$.}
\label{FIG06}
\end{figure}

The guide field, $B_z$ in this paper, can not contribute to the reconnection 
because of the 2.5D approximation. Besides this, the reconnection 
progresses typically with the Alfv\'en time scale, which depends on the 
magnetic field on the $xy$-plane (more exactly speaking, depend on the 
reconnection component of the Alfv\'en velocity). 
It is therefore interesting that how the results change when the 
normalizations are changed: Time is re-normalized by the effective Alfv\'en 
time ($t_\mathrm{A0,eff}=L_0/v_\mathrm{A0,eff}$). 
This means that the same time in Figure \ref{FIG06} is not the same time 
in Figure \ref{FIG09} because the effective Alfv\'en velocity is different 
according to $\theta$. 
The magnetic field is also re-normalized by the initial magnetic field which can reconnect 
($B_0 \sin \theta/2$). When $\theta = 0$, the magnetic field is uniform and the reconnection
does not take place.

\begin{figure}
\begin{center}
\FigureFile(160mm,160mm){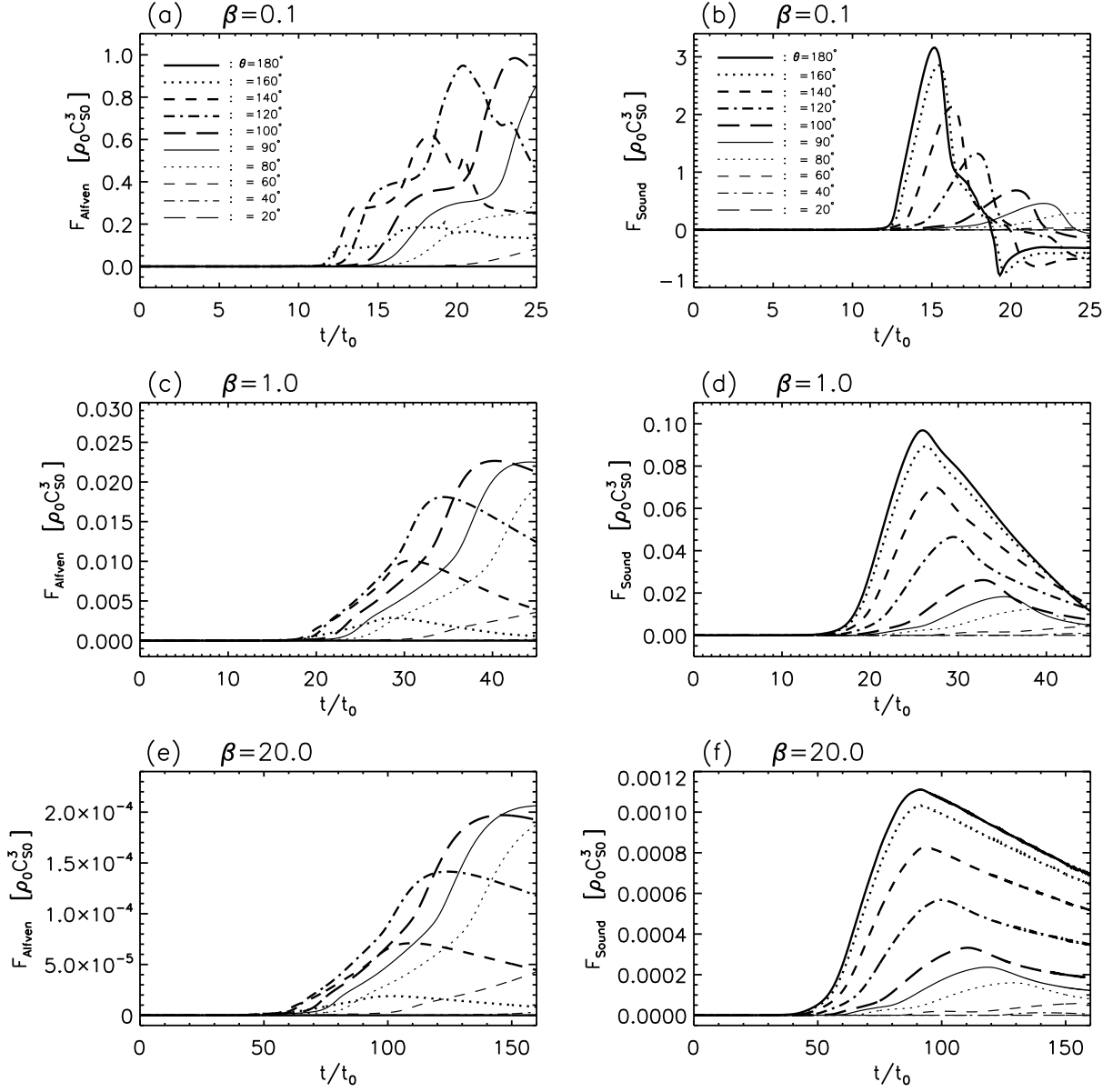}
\end{center}
\caption{Time evolution of the energy fluxes, $F$, (a, c, e) Alfv\'en waves and (b, d, f)
magneto-acoustic waves in each $\theta$ case. 
The plasma $\beta$ are (a, b) 0.1, (c, d) 1.0 and (e, f) 20.0.}
\label{FIG07}
\end{figure}

\begin{figure}
\begin{center}
\FigureFile(160mm,160mm){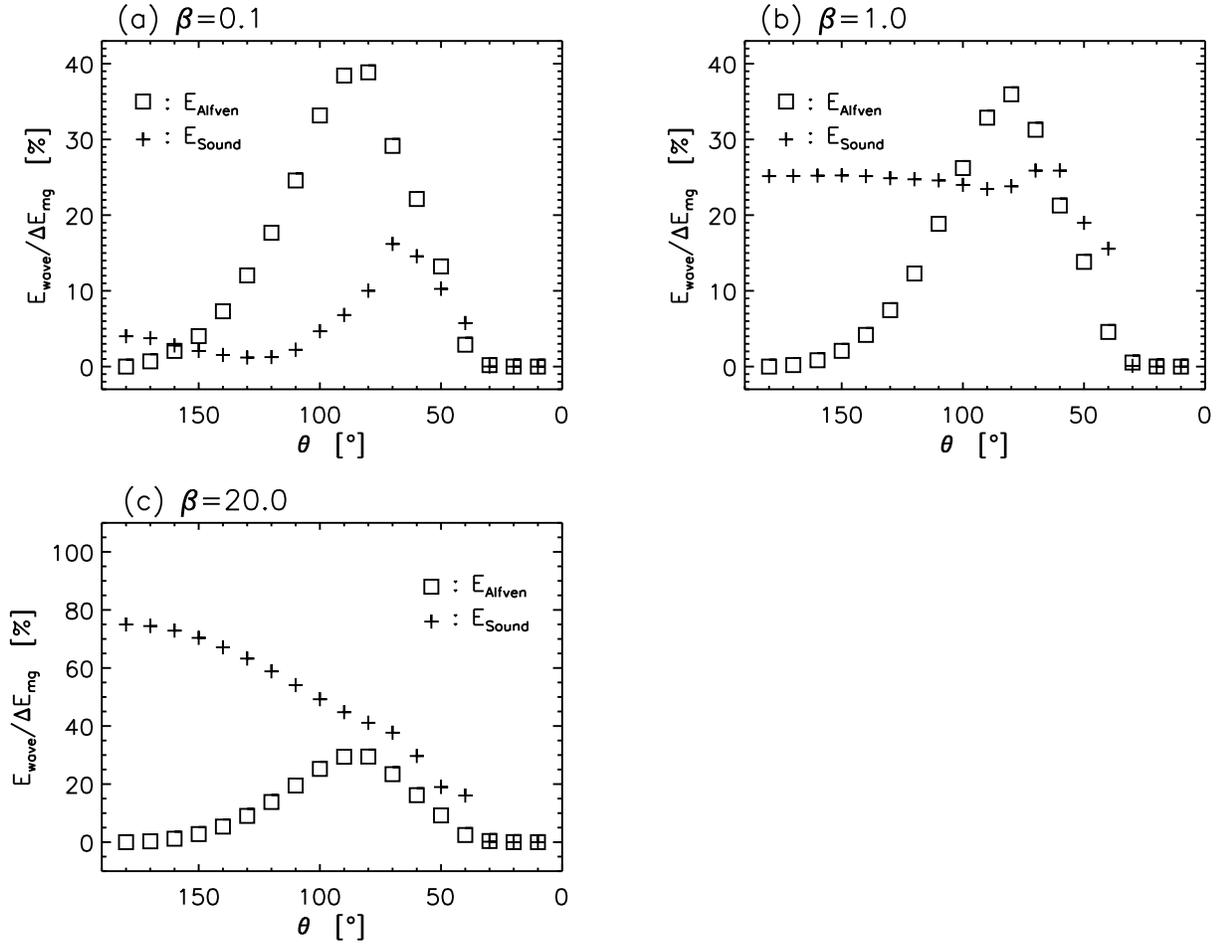}
\end{center}
\caption{The percentage of the energies carried by the Alfv\'en waves
($E_\mathrm{Alfven}/\Delta E_\mathrm{mg}$) and magneto-acoustic waves
($E_\mathrm{Sound}/\Delta E_\mathrm{mg}$) to the released
magnetic energy at the final stage of simulation in the case of $\theta=180^\circ$.}
\label{FIG08}
\end{figure}

\begin{figure}
\begin{center}
\FigureFile(160mm,160mm){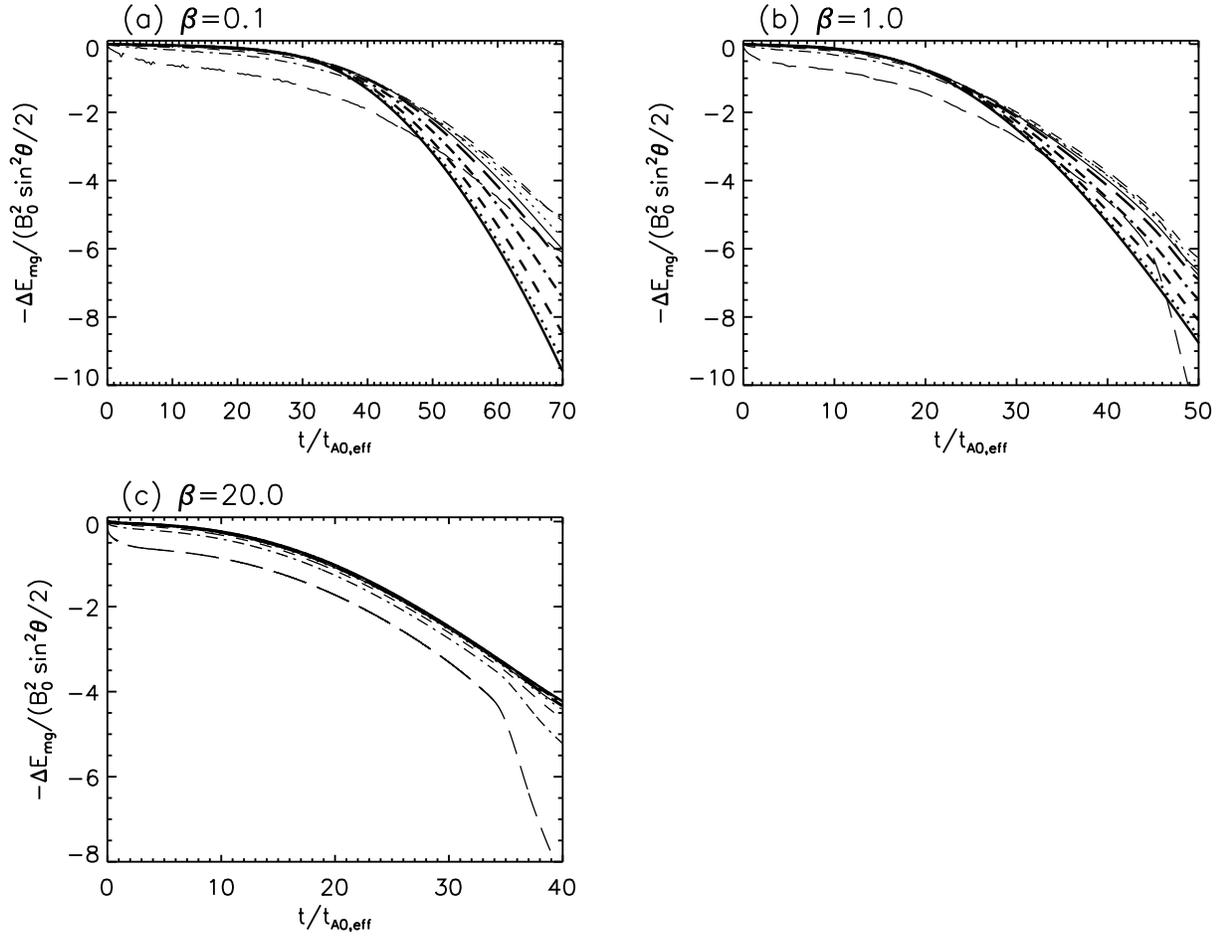}
\end{center}
\caption{Time evolution of the magnetic energy in the computational domain in
each $\theta$ case. Compared with Figure \ref{FIG06}, the horizontal axis, 
time, is re-normalized by the effective Alfv\'en time ($t_\mathrm{A0,eff}=L_0/v_\mathrm{A0,eff}$).
The vertical axis, the difference of the magnetic energy from the initial value, is also 
re-normalized as $-\Delta E_\mathrm{mg}^* = -\Delta E_\mathrm{mg}/({B_0}^2 \sin^2 \theta/2)$.
The line styles are the same as Figure \ref{FIG06}.}
\label{FIG09}
\end{figure}

\begin{figure}
\begin{center}
\FigureFile(160mm,160mm){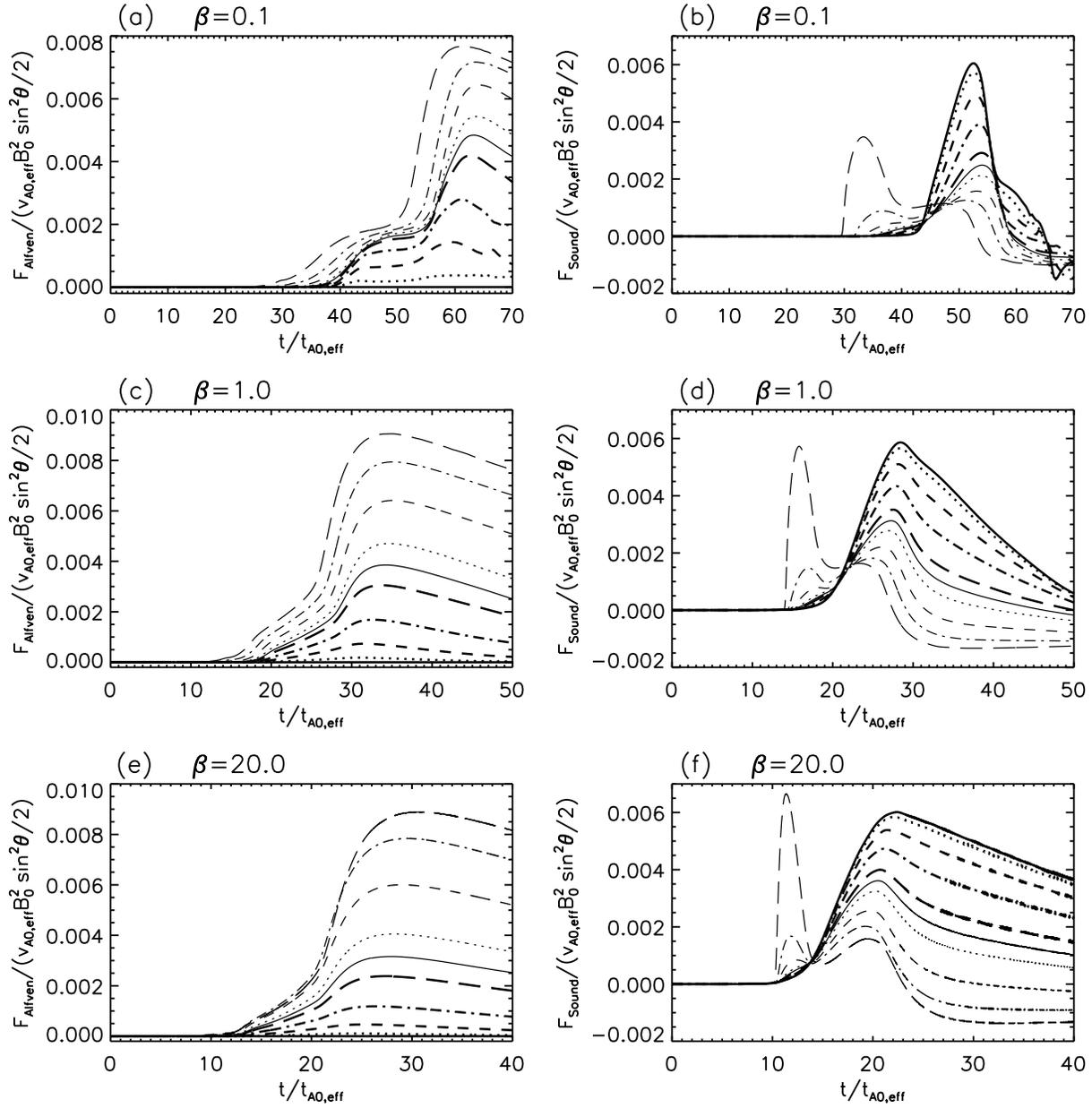}
\end{center}
\caption{Time evolution of the re-normalized energy fluxes, $F^*$, of 
(a, c, e) Alfv\'en waves and (b, d, f) magneto-acoustic waves
in each $\theta$ case. The energy fluxes are
re-normalized as $F^* = F/(v_\mathrm{A0,eff} {B_0}^2 \sin^2 \theta/2)$.
The plasma $\beta$ are (a, b) 0.1, (c, d) 1.0 and (e, f) 20.0.
The line styles are the same as Figure \ref{FIG07}.}
\label{FIG10}
\end{figure}

Figure \ref{FIG09} shows the difference of the re-normalized magnetic energy 
from the initial value as the function of re-normalized time in each $\theta$ 
case. In the high-$\beta$ case, the magnetic energy is released at almost the 
same rate in the effective Alfv\'en time when $\theta \ge 40^\circ$. 
The lower $\beta$ is, the larger the discrepancy of the release rate becomes. 
Figure \ref{FIG10} shows the re-normalized energy fluxes 
and the time integral of those as the function of re-normalized time. The energy flux and its 
time integral of Alfv\'en waves show the time variation similar to each other 
independent of $\beta$. Those of magneto-acoustic waves
also show the time variation similar to each other except when
$\theta$ is relatively small, although there is a bit of $\beta$ dependence.
Compared with Figure \ref{FIG07}, the peak positions are matched.

\begin{figure}
\begin{center}
\FigureFile(160mm,160mm){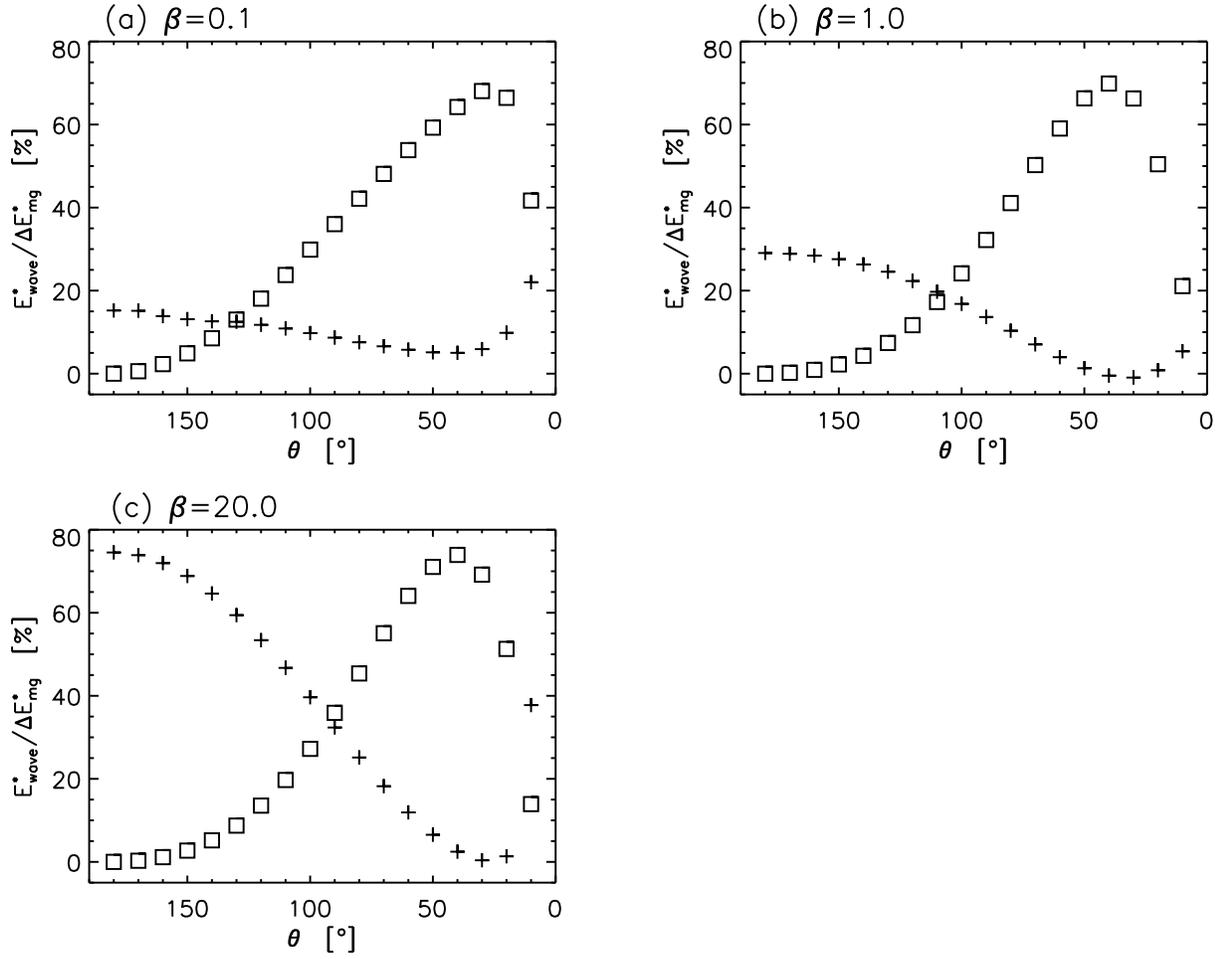}
\end{center}
\caption{The percentage of the energies carried by the Alfv\'en waves
($E_\mathrm{Alfven}^*/\Delta E_\mathrm{mg}^*$) and magneto-acoustic waves
($E_\mathrm{Sound}^*/\Delta E_\mathrm{mg}^*$) to the released
magnetic energy. The released magnetic energy is re-normalized by the energy
of the initial magnetic field which can reconnect.
The symbols are the same as Figure \ref{FIG08}.}
\label{FIG11}
\end{figure}

Figure \ref{FIG11} indicates the percentage of the energies carried by the Alfv\'en
waves and magneto-acoustic waves to the released
magnetic energy. The vertical axis is re-normalized as same as the above-mentioned way.
This shows clearly that the amount of the energy carried by the Alfv\'en waves has almost
the same dependence on $\theta$ independent of $\beta$ {\it if the energy and time are scaled 
by the effective magnetic energy and Alfv\'en time}. This is equivalent to that the
magnetic configuration is important rather than the field strength relative to 
the gas pressure for the energy release rate in the effective Alfv\'en time. 
On the other hand, the amount of the energy carried by the magneto-acoustic waves
shows the $\beta$ dependence.

In this paper, we have reported the results for 2.5D MHD simulations of the magnetic
reconnection. When magnetic fields are non-perfectly antiparallel, the magnetic field 
component perpendicular to the initial field is generated and propagates as the
Alfv\'en wave. We have measured the energy fluxes of Alfv\'en waves and
magneto-acoustic waves. Magneto-acoustic waves
are related to fast mode waves in the high-$\beta$ case ($\beta=20$) and
slow mode waves in the low-$\beta$ case ($\beta=0.1$). The energy carried by the
Alfv\'en waves is more than 30\% of the energy released by the magnetic reconnection
at the maximum. That value satisfies the requirement for energy flux of Alfv\'en waves
necessary for acceleration of high-speed solar wind in the nanoflare coronal heating model.
For more exact discussion, 3-dimensional simulations with more realistic plasma distribution
are necessary.

We would like to thank David H. Brooks and Takeru Suzuki for helpful comments. 
This study was initiated as a part of the ACT-JST summer school for numerical simulations
in astrophysics and in space plasmas. Numerical computations were carried out on VPP5000 
(project ID: yhk32b, rhk05b, and whk08b) and the general-purpose PC farm (project P.I. KT)
at Center for Computational Astrophysics, CfCA, of National Astronomical Observatory of Japan.
This work was supported by the Grant-in-Aid for the 21st Century COE "Center for Diversity
and Universality in Physics" from the Ministry of Education, Culture, Sports, Science and
Technology (MEXT) of Japan.



\begin{table}[htpb]
\caption{Values of Parameters in Simulations}
\begin{center}
\begin{tabular}{c|c|c}
\hline \hline
plasma $\beta$ & Alfv\'en velocity, $v_\mathrm{A0}$ & $R_\mathrm{m}$ \\ \hline
 0.1 & $\sim 3.5$ & $\sim 2500$\\ 
 1 & $\sim  1.1$ & $\sim  800$\\
20 & $\sim 0.24$ & $\sim  180$\\
\hline
\end{tabular}
\end{center}
\end{table}

\clearpage

\clearpage

\clearpage

\clearpage

\clearpage

\clearpage

\end{document}